\documentclass[12pt]{iopart}
\usepackage{iopams}  
\usepackage{setstack}
\usepackage{graphicx}
\usepackage{float}
\usepackage[colorlinks = true]{hyperref}

\newcommand{\bp}{\mathbf{p}}
\newcommand{\bA}{\mathbf{A}}
\newcommand{\bB}{\mathbf{B}}
\newcommand{\bC}{\mathbf{C}}
\newcommand{\bH}{\mathbf{H}}

\begin{document}

\title{Tightest bound on hidden entropy production from partially observed dynamics}

\author{Jannik Ehrich}

\address{Department of Physics, Simon Fraser University, Burnaby, British Columbia, V5A 1S6, Canada\\
Institut f\"ur Physik, Carl von Ossietzky Universit\"at, 26111 Oldenburg, Germany}
\ead{jehrich@sfu.ca}
\vspace{5mm}
Version: \today

\begin{abstract}
Stochastic thermodynamics allows us to define heat and work for microscopic systems far from thermodynamic equilibrium, based on observations of their stochastic dynamics. However, a complete account of the energetics necessitates that all relevant nonequilibrium degrees of freedom are resolved, which is not feasible in many experimental situations. A simple approach is to map the visible dynamics onto a Markov model, which produces a lower-bound estimate of the entropy production. The bound, however, can be quite loose, especially when the visible dynamics only have small or vanishing observable currents. An alternative approach is presented that uses all observable data to find an underlying hidden Markov model responsible for generating the observed non-Markovian dynamics. For masked Markovian kinetic networks, one obtains the tightest possible lower bound on entropy production of the full dynamics that is compatible with the observable data. The formalism is illustrated with a simple example system.
\end{abstract}

%\submitto{\JSTAT}
\maketitle

\section{Introduction}
% Background
Small-scale systems are heavily influenced by thermal fluctuations, making it a hard task to estimate their energetics. Remarkably, a quantification of a system's dissipation is possible by observation of time-irreversibility of its stochastic dynamics. Within the framework of stochastic thermodynamics~\cite{Jarzynski2011,Seifert2012,VandenBroeck2015}, one assigns entropy production to individual trajectories, as well as ensemble averages quantifying a system's average rate of entropy production.

% Motivation
However, a full thermodynamic accounting relies on a thorough separation of time scales which dictates that fast equilibrated dynamical variables belong to the \emph{heat bath} while slow ones constitute the \emph{system} and must be observed to infer the correct dissipation. In many experimental situations this is not the case, e.g., when observing molecular motors where often the trajectories of an attached cargo are available while the motor position and its internal chemical reactions remain hidden~\cite{Kolomeisky2013}.

It is therefore important to study how one can properly account for such \emph{hidden slow degrees of freedom}. Here, I focus on ways to estimate the average rate of entropy production when not all degrees of freedom of a system are visible.

% State-of-the-art
A collection of results showing how thermodynamic consistency can lead to information about hidden system properties goes under the name of \emph{thermodynamic inference}~\cite{Seifert2019}. It is usually assumed that there exists an underlying Markov process describing the dynamics of all relevant degrees of freedom. An observer only has access to (measurements of) a subset of these, thus rendering the observed process non-Markovian.

A first approximation that yields estimates of entropy production is to completely ignore the non-Markovian nature of the observed trajectories. This approach might be chosen naturally by not being aware that there are unobserved slow degrees of freedom---it is, after all, not always easy to realize that a given time-series has non-Markovian statistics. Mapping the coarse-grained system to a Markov process results in a model that produces the correct state probabilities of the coarse-grained states and transition rates between them. Generally, this procedure produces a lower bound on the real average entropy production rate~\cite{Esposito2012} which becomes tight when the hidden dynamics obey detailed balance and evolve on much shorter time-scales than the observed ones~\cite{Bo2014}, thus effectively rendering them bath variables.

Importantly, whenever there are no observable probability currents, mapping to a Markov process gives a model that has zero entropy production~\cite{Martinez2019} even though the underlying process may be out of equilibrium. Moreover, the estimates of trajectory entropy of the Markov process do not fulfill the usual fluctuation theorems~\cite{Rahav2007,Mehl2012,Chiang2016,Uhl2018,Kahlen2018} which can perhaps be used to detect that a Markovian description is inadequate.

A related approach that potentially requires even less information about the system is to use the so-called \emph{thermodynamic uncertainty relation}~\cite{Horowitz2020,Barato2015,Gingrich2016,Gingrich2017,Proesmans2017} which from the fluctuations of observed currents produces a lower-bound estimate of the entropy production. Under ideal conditions, one may find a tight bound from the fluctuations of an optimal current without detailed knowledge of the underlying system dynamics~\cite{Li2019,Manikandan2020,Vu2020,Otsubo2020}.

When more detailed information about the statistics of the observed trajectories is available, time-irreversibility can be used to estimate entropy production. This approach relies on the observation that the time-irreversibility of the full system trajectories (as quantified by the relative entropy between trajectories forwards and backwards in time) is equal to the entropy production~\cite{Kawai2007,Horowitz2009,Parrondo2009}. When the observed trajectories are coarse-grained in time, phase space, or both, their time-irreversibility yields a lower bound on the complete entropy production rate~\cite{Gomez-Marin2008,Roldan2010,Roldan2012,Muy2013}. 

Here, I will focus on a coarse-graining in state space which lumps together several microstates into observable meso-states~\cite{Esposito2012,Bo2014}. Specifically, I will further assume that the state space is \emph{masked}, such that the observer only sees a subset of transitions between a few microstates. All other microstates are lumped together into one hidden mesostate. For this setting, different inference strategies~\cite{Shiraishi2015,Polettini2017,Bisker2017} for the average rate of entropy production have been proposed that produce lower-bound estimates comparable to mapping the observed dynamics onto a Markov model.

Recently, Mart\'{i}nez \emph{et al.}~\cite{Martinez2019} showed how information about the waiting-time distribution of a masked Markov jump process can be leveraged to estimate the full time-series irreversibility of the observed trajectories. This result is remarkable, as it allows the estimation of full time-series irreversibility from the sampling of waiting time-distributions instead of whole trajectories, which is more feasible in an experiment.

% Knowledge gaps
However, the lower-bound estimates inferred from time-series irreversibility of observed trajectories are still often orders of magnitude off from the real rate of entropy production, which prompts the question of whether one can do better, e.g., by fitting the underlying Markov process to the observable data. This possibility was also mentioned in passing in a recent letter by Teza and Stella~\cite{Teza2020} on a related subject. Here, I illustrate how this task can be accomplished by considering a simple model and discrete-time Markov-chain dynamics.

% Objectives
% 1) calculate time-series irreversibility for masked Markov chains
% 2) Fitting hidden transition matrix to observable statistics
% 3) Infer bounds on entropy production from possible hidden transition matrices
The aim of this paper is to (1) show how to efficiently calculate the time-series irreversibility for such dynamics; (2) from this, illustrate how to constrain the hidden transition matrix to match the observable statistics for a simple model; and (3) use the remaining freedom in the hidden transition probabilities to find possible entropy production rates of the underlying full dynamics compatible with the observable trajectories, which leads to an improved lower bound estimate of the real dissipation.

\section{Masked Markov chains}
Consider a stochastic system modeled by a discrete-time Markov chain $z_t$ on a finite set of connected states labeled $1$ through $K$. The probability of the system being in each state at time $t$ is given by the vector $\mathbf{p}_t$. Its evolution is described by the master equation
\begin{equation}
\bp_{t+1} = \bA\,\bp_{t},
\end{equation}
where $\bA$ is the $K\times K$ \emph{transition probability matrix}. Its entry $a_{ij}$ in row $i$ and column $j$ gives the probability $p(z_{t+1}=i|z_t=j)$ to go from state $j$ to state $i$ in one time step. Due to probability conservation, the columns of $\bA$ sum to $1$.

Assuming that the system is in the \emph{steady state}, the stationary probabilities $\bp$ are given by the solution of
\begin{equation}
\bp = \bA\, \bp.
\end{equation}
Additionally, it is demanded that every transition is \emph{not absolutely irreversible}~\cite{Murashita2014}, such that $a_{ij} \neq 0 \Leftrightarrow a_{ji} \neq 0$.

The average \emph{steady-state entropy production} is defined as~\cite{Seifert2012,VandenBroeck2015}
\begin{equation}\label{eq:entropy_prod}
\Delta \Sigma = \sum\limits_{i,j} a_{ij} p_j \ln\frac{a_{ij} p_j}{a_{ji} p_i} \geq 0.
\end{equation}
For thermodynamic systems, this entropy production is proportional to the average dissipated energy. If it is nonzero in the stationary regime, the system is in a \emph{nonequilibrium steady state}. Then, energy is spent as \emph{housekeeping heat} to maintain steady-state probability currents~\cite{Seifert2012}.

\begin{figure}[htpb]
    \centering
    \includegraphics[width = 0.55 \linewidth]{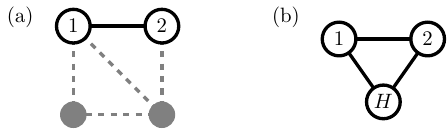}
    \caption{Example masked network of states. (a) Network of states of the complete process. The labeled states 1 and 2 are observed. All other states (grey) are not observed. (b) Observed network of states. All unobserved states are lumped together into one \emph{hidden} state $H$.}
    \label{fig:schematic_hidden_network}
\end{figure}

\begin{figure}[htpb]
    \centering
    \includegraphics[width = 0.7 \linewidth]{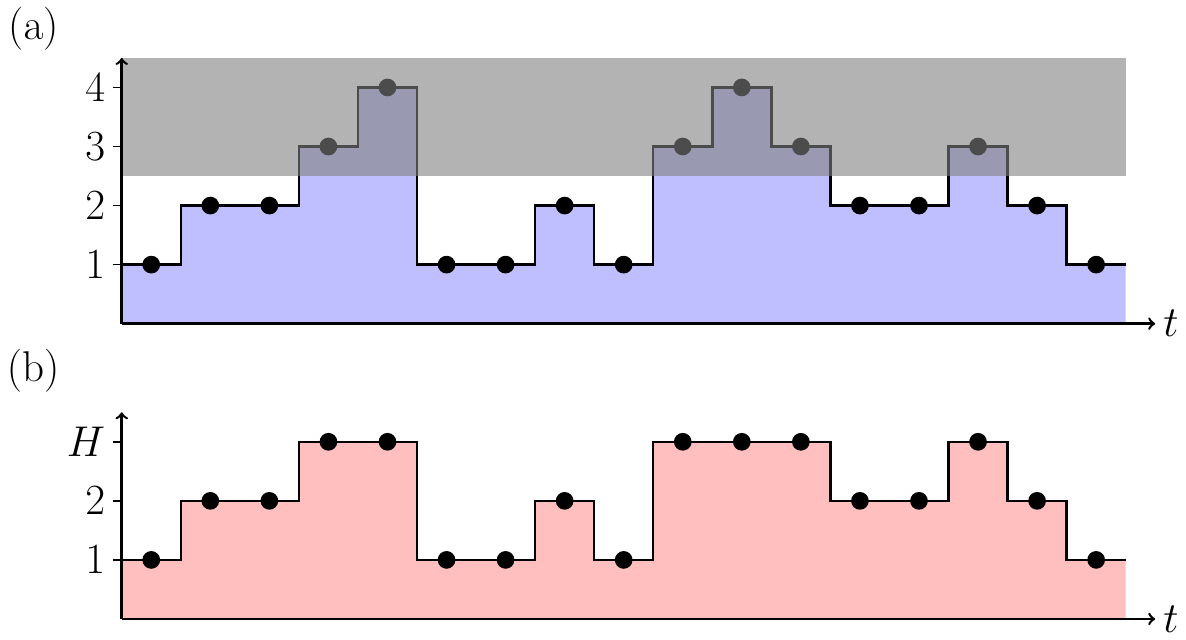}
    \caption{Sample trajectory from a masked Markov jump process: (a) Full trajectory and (b) observed trajectory.}
    \label{fig:sample_trajectory}
\end{figure}

From~\eref{eq:entropy_prod} it is evident that to calculate the entropy production, one needs to be able to monitor every state and every transition in the network of states. When some states are \emph{masked}, an observer is forced to conclude that whenever they do not see the process being in one of the visible states, it has to be in the hidden state $H$. Therefore, from the point of view of an observer, all unobserved states are lumped together. Figure~\ref{fig:schematic_hidden_network} depicts a schematic setup. The network of states is shown as a graph whose vertices are the states and whose edges indicate transitions (in both directions) between these states. A sample trajectory is depicted in figure~\ref{fig:sample_trajectory}.

\begin{figure}[htpb]
    \centering
    \includegraphics[width = 0.55 \linewidth]{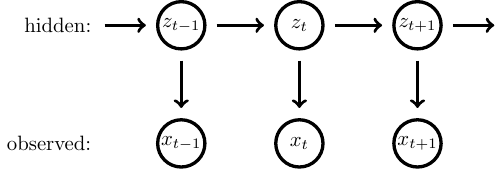}
    \caption{Causal diagram of a hidden Markov model. At each time step the Markov chain $z_t$ is hidden and only observations $x_t$ are available.}
    \label{fig:HMM_causal_diagram}
\end{figure}

The complete process has the structure of a \emph{hidden Markov chain}, which consists of a Markov chain $z_{1:N} := \{z_1,z_2,...,z_N\}$ of hidden states that are correlated with observations $x_{1:N} := \{x_1,x_2,...,x_N\}$. The observations themselves are assumed to be \emph{memoryless}: at each time step they solely depend on the current hidden state and not on the past measurement history. This causal structure is visualized in the \emph{causal diagram} in figure~\ref{fig:HMM_causal_diagram} where time runs horizontally and causal influences between different states are marked by arrows. 

\section{Bounds on the rate of entropy production}
Having established the setup, I will present the two methods for estimating the entropy production from the partial information available. It is assumed that a long trajectory of observations is available from which one can estimate statistical properties of the system (by virtue of the ergodicity of the underlying Markov chain).

\subsection{Mapping to a Markov process}
The procedure of mapping the visible dynamics to a Markov model is not unique and different coarse-graining schemes have been proposed~\cite{Esposito2012,Bo2014,Polettini2017,Teza2020,Puglisi2010,Altaner2012,Zimmermann2015}. However, a natural choice is that model which one would naively infer from the visible dynamics if unaware of their non-Markovian nature.

For the example here the procedure amounts to estimating the relative frequency of transitions by counting jumps between states $1$, $2$, and $H$ in a long trajectory. Given the counts of jumps between states $i$ and $j$, $n[j \to i]$, the \emph{coarse-grained} jump probability $\tilde a_{ij}$ is
\begin{equation}
    \tilde a_{ij} = \frac{n[j \to i]}{\sum\limits_i n[j \to i]}\,.
\end{equation}

Together with the observed coarse-grained steady-state probabilities, $\tilde p_i$, for $i~\in~\{1,2,H\}$, one obtains the average rate of \emph{coarse-grained} entropy production~\cite{Esposito2012, Bo2014}, sometimes also called \emph{apparent} entropy production~\cite{Mehl2012,Uhl2018,Kahlen2018}:
\begin{equation}
    \Delta \tilde\Sigma := \sum\limits_{i , j \in \{1,2,H\}} \tilde a_{ij} \tilde p_j \ln\frac{\tilde a_{ij} \tilde p_j}{\tilde a_{ji} \tilde p_i}\,. \label{eq:coarse-grained_EP}
\end{equation}

\subsection{Time-series irreversibility} \label{sec:time_series_irreversibility}
First, I show that the entropy production inferred from the time-series irreversibility of the observed data underestimates the true entropy production. The entropy production per time step in~\eref{eq:entropy_prod} can be written both as the time-series irreversibility of a long trajectory of the full process (see, e.g., Ref.~\cite{Roldan2012}) as well as the joint time-series irreversibility of the visible \emph{and} the full trajectory:
\begin{eqnarray}
\Delta \Sigma = \Delta \Sigma_{\mathrm{DKL}}[z_{1:N}] &:= \lim\limits_{N \rightarrow \infty} \frac{1}{N} \sum\limits_{z_{1:N}} p(z_{1:N}) \ln\frac{p(z_{1:N})}{p(\bar z_{1:N})}\\
&=\lim\limits_{N \rightarrow \infty} \frac{1}{N} \sum\limits_{z_{1:N}} p(z_{1:N}) \ln\frac{p(z_{1:N})\, p(x_{1:N}|z_{1:N})}{p(\bar z_{1:N})\,p(\bar x_{1:N}|\bar z_{1:N})}\\
&=\lim\limits_{N \rightarrow \infty} \frac{1}{N} \sum\limits_{x_{1:N},z_{1:N}} p(x_{1:N},z_{1:N}) \ln\frac{p(x_{1:N},z_{1:N})}{p(\bar x_{1:N},\bar z_{1:N})} \\
&=: \Delta \Sigma_{\mathrm{DKL}}[x_{1:N},z_{1:N}],
\end{eqnarray}
where $\bar z_{1:N} = \{z_N, z_{N-1},...,z_1 \}$ and similarly for $\bar x_{1:N}$. The second line follows from the fact that given the trajectory of the true states, the observations are uncorrelated (i.e., memoryless) in the original process and its time-reverse:
\begin{equation}
    \hspace{-1cm} p(x_{1:N}|z_{1:N}) = p(x_1,x_2,...x_N|z_1,z_2,...z_N) = \prod_{i=1}^N p(x_i|z_i) = \bar p(\bar x_{1:N}|\bar z_{1:N}).
\end{equation}
In fact, these conditional probabilities are the same products of Kronecker-deltas for the masked dynamics considered here.

The time-series irreversibility of the observed trajectory alone captures less irreversibility:
\begin{eqnarray}
\Delta \Sigma_{\mathrm{DKL}}[x_{1:N},z_{1:N}] &= \lim\limits_{N \rightarrow \infty} \frac{1}{N} \sum\limits_{x_{1:N}} p(x_{1:N}) \ln\frac{p(x_{1:N})}{p(\bar x_{1:N})}\nonumber\\
&\qquad\qquad\times\left[1 + \sum\limits_{z_{1:N}} p(z_{1:N}|x_{1:N}) \ln\frac{p(z_{1:N}|x_{1:N})}{p(\bar z_{1:N}|\bar x_{1:N})} \right] \\ 
 &\geq \Delta\Sigma_{\mathrm{DKL}}[x_{1:N}].
\end{eqnarray}

Thus, $\Delta\Sigma_{\mathrm{DKL}}[x_{1:N}] \leq \Delta\Sigma_{\mathrm{DKL}}[x_{1:N},z_{1:N}] = \Delta\Sigma_{\mathrm{DKL}}[z_{1:N}] = \Delta\Sigma$, which makes sense since the observations alone contain less information about time-irreversibility than the full sequence, and once the full sequence is known, the observations are redundant information. The time-series irreversibility of the observable trajectories therefore serves as another lower bound for the complete entropy production.

However, estimating time-series irreversibility from the frequency of occurrence of different trajectories is no easy task. One approach in this setting would be to estimate it by counting the frequency of \emph{substrings} of increasing length in a long trajectory and using clever extrapolation techniques~\cite{Roldan2010, Roldan2012}.

A novel approach was presented recently by Mart\'{i}nez \emph{et al.}~\cite{Martinez2019} who realized that the observed dynamics form a \emph{semi-Markov} process, i.e., a Markov jump process with non-Poissonian waiting times. This observation allows one to calculate time-series irreversibility of a stationary masked jump process from the relative frequency of jumps \emph{and} waiting times conditioned on jumps. In the following, I show how their strategy can be adapted for discrete-time Markov chains.

Notice that the visible trajectory in figure~\ref{fig:sample_trajectory}(b) is fully characterized by the sequence of jumps between visible states and the number of time steps needed to complete these jumps. A trajectory of length $N$ with $M$ jumps is thus specified by
\begin{equation}
    x_{1:N} = \left\{ (x_1 \to x_2,n_1), (x_2 \to x_3,n_2), ..., (x_M \to x_{M+1},n_M) \right\}\,,
\end{equation}
where $x_i \in \{1,2\}$ indicate the visible states and $n_i$ the number of time steps per jump between visible states, such that $\sum_{i=1}^M n_i = N-1$.

\begin{figure}[htpb]
    \centering
    \includegraphics[width = 0.8 \linewidth]{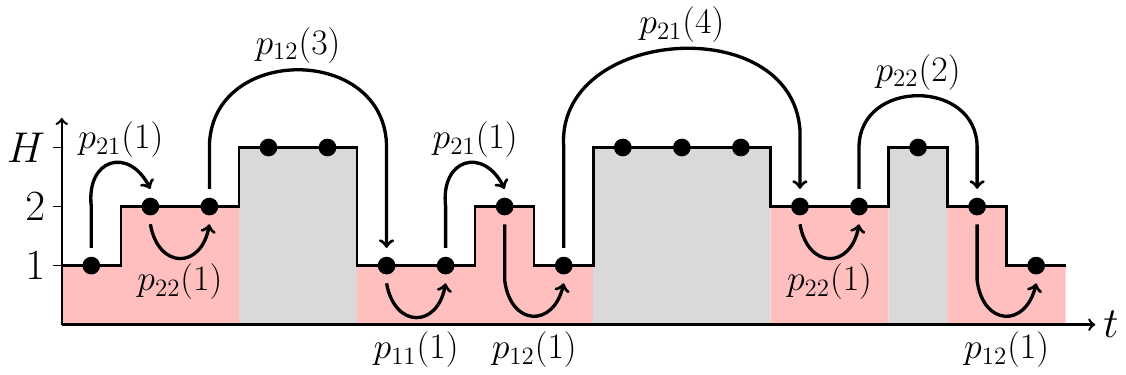}
    \caption{Visible trajectory annotated with jump probabilities between visible states. The functions $p_{ij}(n)$ encode the probability to jump from visible state $j$ to visible state $i$ in $n$ time steps.}
    \label{fig:trajectory_annotated}
\end{figure}

Let $p_{ij}(n)$ be the probability to jump from visible state $j$ to visible state $i$ in $n$ time steps, i.e., with $n-1$ intermediate time steps spent in the hidden part of the network. Figure~\ref{fig:trajectory_annotated} gives an example by indicating the relevant probabilities in the example trajectory in figure~\ref{fig:sample_trajectory}(b).

The probability of visible trajectory $x_{1:N}$ is thus
\begin{equation}
    p(x_{1:N}) = \pi_{x_1} \, p_{x_2 x_1}(n_1) \, p_{x_3 x_2}(n_2)\,...\, p_{x_{M+1} x_M}(n_M)\,, \label{eq:forward_traj_prob}
\end{equation}
where
\begin{equation}
    \pi_i := \frac{p_i}{p_1 + p_2} \label{eq:steady-state_probs_cond}
\end{equation}
is the steady-state probability to find the process in state $i$ given it is in one of the visible states. Note that this restricts the set of all possible trajectories to those that start and end in the visible part of the network, which can be accomplished by ignoring the first and last hidden jumps in a long trajectory.

The probability of the time-reversed trajectory reads
\begin{equation}
    p(\bar x_{1:N}) = \pi_{x_{M+1}} \, p_{x_M x_{M+1}}(n_M)\,...\, p_{x_2 x_3}(n_2)\,...\, p_{x_{1} x_2}(n_1)\,. \label{eq:reverse_traj_prob}
\end{equation}
Then the time-irreversibility per jump between visible states is:
\begin{eqnarray}
\hspace{-1.7cm}\delta \Sigma_\mathrm{DKL}[x_{1:N}] &= \lim\limits_{M\rightarrow \infty} \frac{1}{M}\sum\limits_{x_1,...,x_{M+1}} \sum\limits_{n_1,...,n_M} p(x_{1:N})\left[\ln\frac{\pi_{x_1}}{\pi_{x_{M+1}}} + \ln \prod\limits_{k=1}^M \frac{p_{x_{k+1} x_k}(n_k)}{p_{x_{k} x_{k+1}}(n_k)} \right]\\
    &= \lim\limits_{M\rightarrow \infty} \frac{1}{M}\sum\limits_{x_1,...,x_{M+1}} \sum\limits_{n_1,...,n_M} p(x_{1:N})\left[ \ln \prod\limits_{k=1}^M \frac{p_{x_{k+1} x_k}(n_k)\,\pi_{x_k}}{p_{x_{k} x_{k+1}}(n_k) \,\pi_{x_{k+1}}} \right]\\
    &= \lim\limits_{M\rightarrow \infty} \frac{1}{M}\sum\limits_{k=1}^M \sum\limits_{x_k,x_{k+1}} \sum\limits_{n_k} p_{x_{k+1} x_k}(n_k) \pi_{x_k} \ln\frac{p_{x_{k+1} x_k}(n_k)\,\pi_{x_k}}{p_{x_{k} x_{k+1}}(n_k) \,\pi_{x_{k+1}}} \label{eq:DKL_per_jump_2}\\
    &= \sum\limits_{i,j=1}^2 \sum\limits_{n=1}^\infty p_{ij}(n) \pi_j\ln\frac{p_{ij}(n)\,\pi_j}{p_{ji}(n)\,\pi_i}\,. \label{eq:DKL_per_jump}
\end{eqnarray}

To get the irreversibility per time step, $\delta\Sigma_\mathrm{DKL}$ needs to be divided by the average number $\bar n$ of time steps per jump between the visible states. In a long trajectory this number is equal to the total number $N$ of symbols divided by the number of jumps, which is $N-n_H$, where $n_H$ is the total number of hidden symbols. Thus,
\begin{eqnarray}
\bar n &= \lim_{N\to\infty} \frac{N}{N - n_H} = \lim_{N\to\infty} \frac{1}{1 - n_H/N} = \frac{1}{1-\tilde p_H} = \frac{1}{p_1 + p_2}\,,
\end{eqnarray}
where $\tilde p_H$ denotes the steady-state probability to observe a hidden symbol.

Therefore, with~\eref{eq:steady-state_probs_cond}~and~\eref{eq:DKL_per_jump}, the time-series irreversibility per time step reads
\begin{eqnarray}
\Delta \Sigma_\mathrm{DKL}[x_{1:N}] &= \frac{\delta \Sigma_\mathrm{DKL}(x_{1:N})}{\bar n}\\
    &= \sum\limits_{i,j=1}^2 \sum\limits_{n=1}^\infty p_{ij}(n)p_j \ln\frac{p_{ij}(n)\,p_i}{p_{ji}(n)\,p_j}\\ 
    &= \sum\limits_{n=1}^\infty \left[ p_{12}(n)\,p_2 - p_{21}(n)\,p_1 \right]\ln\frac{p_{12}(n)\,p_2}{p_{21}(n)\,p_1}\,, \label{eq:DKL_per_step}
\end{eqnarray}
which can now be calculated from observations of the jump probabilities $p_{ij}(n)$. Conveniently, since every term is nonnegative, the sum can be truncated at any $n$ and still yield a lower-bound estimate of entropy production.

\subsection{Example} \label{sec:example_network}
For an explicit example, consider a network of four states with the topology depicted in figure~\ref{fig:schematic_hidden_network}, specified by the transition matrix
\begin{equation}
    \bA = \left( \begin{array}{cccc}
0.4 - 0.1\,e^{\Delta\mu/2} & 0.2\,e^{-\Delta\mu/2} & 0.3 & 0.3 \\
0.1\,e^{\Delta\mu/2} & 0.9 - 0.2\,e^{-\Delta\mu/2} & 0.1 & 0\\
0.1 & 0.1 & 0.4 & 0.6\\
0.5 & 0 & 0.2 & 0.1\end{array} \right)\,, \label{eq:example_transition_matrix}
\end{equation}
where $\Delta\mu$ controls the transition probabilities at the edge $1 -2$. In particular, for $\Delta\mu \approx 0.86$, there is no net current flowing over the visible edge.

From the underlying transition rates, the steady-state probabilities $p_i$ can be calculated which also give the coarse-grained probabilities $\tilde p_i$:
\begin{equation}
    \tilde p_i=\cases{
    p_i & for $i \leq 2$\\
    \sum\limits_{j > 2} p_j & for $i = H$\\}\,. \label{eq:coarse-grained_steady_state_prob}
\end{equation}

The coarse-grained transition probabilities $\tilde a_{ij}$ are calculated as follows~\cite{Esposito2012}:
\begin{equation}
    \tilde{a}_{ij} = \cases{
        a_{ij} & for $i,j \leq 2$\\
        a_{3 j} + a_{4j} & for $i= H, j \leq 2$\\
        \frac{a_{i3}\,p_3 + a_{i4}\,p_4}{p_3+p_4} & for $i \leq 2 , j = H$\\
        \frac{(a_{33}+a_{43})\,p_3 + (a_{34}+a_{44})\,p_4}{p_3+p_4} & for $i,j = H $
    }\,. \label{eq:calc_coarse-grained_matrix}
\end{equation}
These probabilities form a transition matrix for the Markovian dynamics on the reduced network of states. They can be used in~\eref{eq:coarse-grained_EP} to calculate the coarse-grained entropy production.

To calculate the jump probabilities $p_{ij}(n)$, note first that in general the transition matrix $A$ can be written in block form,
\begin{equation}
    \bA = \left( \begin{array}{cc}
    \left( \begin{array}{cc}
        a_{11} & a_{12}\\
        a_{21} & a_{22}
\end{array} \right) & \bC \\
    \bB & \bH
\end{array} \right)\,, \label{eq:block_matrix}
\end{equation}
where $\bB$ denotes the matrix of probabilities of transitions from the visible part into the hidden part of the network, $\bC$ labels those transition probabilities from the hidden to the visible parts, and $\bH$ denotes transition probabilities between the hidden states. The probabilities for the jumps between the visible states constitute the jump probabilities with length $1$, $p_{ij}(1)$, and can therefore be directly observed from the time-series. The other sub-matrices only appear as products in the observable jump probabilities.

\begin{figure}[ht]
    \centering
    \includegraphics[width = 0.6 \linewidth]{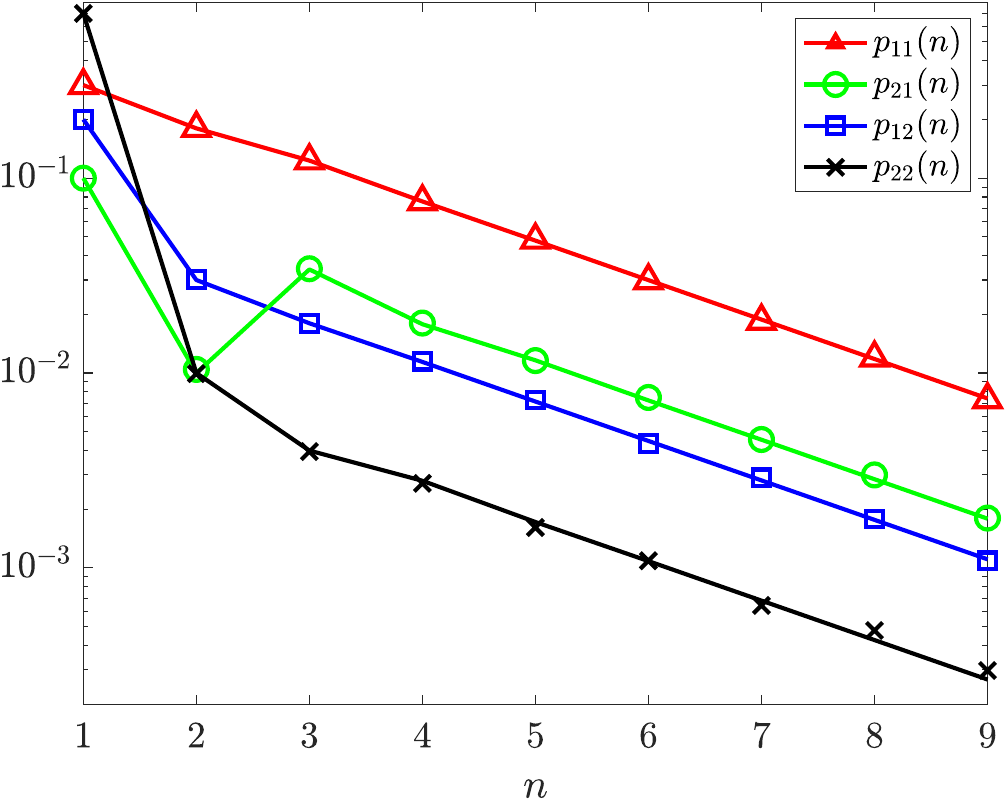}
    \caption{Jump probabilities $p_{ij}(n)$ for the example network specified by the transition probability matrix in~\eref{eq:example_transition_matrix} for $\Delta\mu=0$. Solid lines are analytical calculations using~\eref{eq:jump_probabilities} and symbols are estimations using a simulated trajectory of length $T=10^6$.}
    \label{fig:jump_probabilities}
\end{figure}

To jump from one visible state to another in $n\geq 2$ time steps, the process needs to jump from the visible into the hidden part of the network, then jump $n-2$ times among the hidden states, and finally jump into the visible part. Therefore, the jump probabilities $p_{ij}(n)$ are given by
\begin{equation}
    p_{ij}(n) = \cases{
    a_{ij} & for $n = 1$\\
    \left[ \bC \, {\bH}^{n-2}\, \bB \right]_{ij} & for $n \geq 2$
    }\,. \label{eq:jump_probabilities}
\end{equation}
Figure~\ref{fig:jump_probabilities} shows these jump probabilities for the first few $n$.

\begin{figure}[ht]
    \centering
    \includegraphics[width = 0.6 \linewidth]{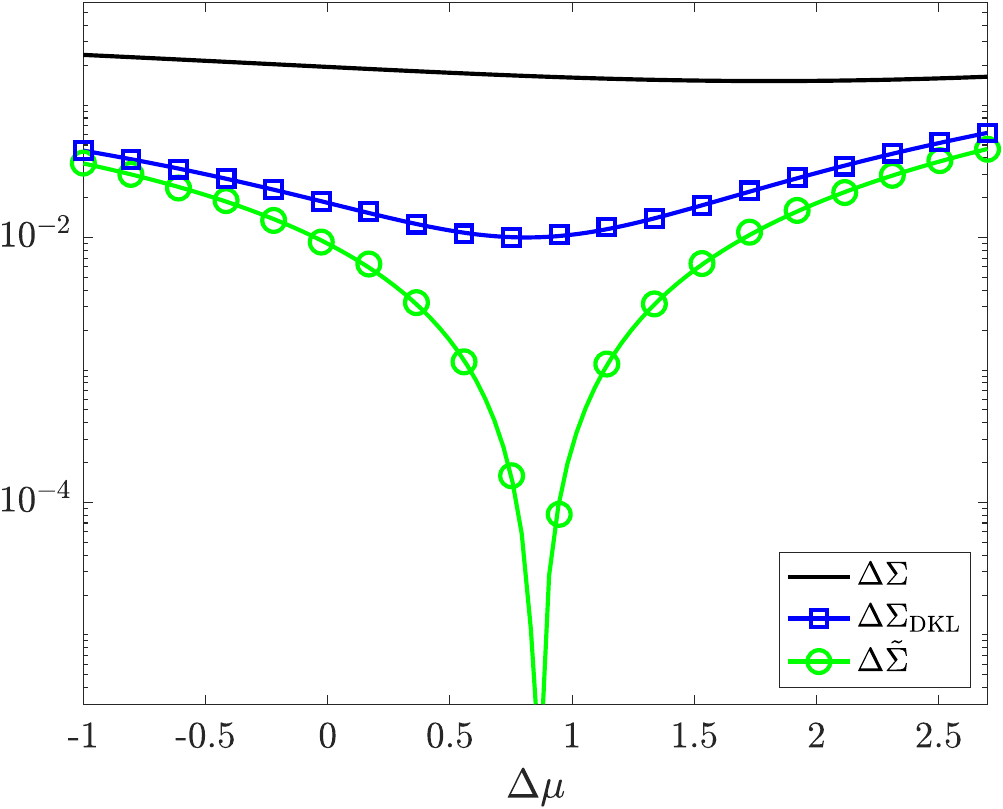}
    \caption{Real entropy production $\Delta \Sigma$~\eref{eq:entropy_prod} per time step, coarse-grained entropy production $\Delta\tilde\Sigma$~\eref{eq:coarse-grained_EP} resulting from mapping to a Markov model, and observable time-series irreversibility $\Delta\Sigma_\mathrm{DKL}$~\eref{eq:DKL_per_step}, as a function of $\Delta\mu$. Solid lines are the entropy productions using analytical calculations of the transition and jump probabilities. Symbols are estimations of transition and jump probabilities from a simulated trajectory of length $T=10^6$. In the absence of observable currents at $\Delta\mu \approx 0.86$ $\Delta\tilde\Sigma$ indicates a reversible process while $\Delta\Sigma_\mathrm{DKL}$ still captures some irreversibility.}
    \label{fig:example_EP_1}
\end{figure}

Having determined the coarse-grained transition probabilities $\tilde a_{ij}$ as well as the jump probabilities $p_{ij}(n)$, one can calculate the coarse-grained entropy production per time step $\Delta \tilde\Sigma$ and the time-series irreversibility $\Delta \Sigma_\mathbf{DKL}$ from~\eref{eq:coarse-grained_EP}~and~\eref{eq:DKL_per_step}, respectively. Figure~\ref{fig:example_EP_1} shows the results of this procedure alongside the real entropy production of the underlying process. For $\Delta\mu \approx 0.86$ the coarse-grained entropy production indicates a reversible process because there are no net currents among the observed states $1$, $2$, and $H$.

One sees that time-series irreversibility is generally a tighter bound on the real entropy production and captures at least some of the irreversibility in the absence of observable currents. However, even the time-series irreversibility underestimates the real entropy production, sometimes by more than an order of magnitude. It is therefore natural to ask whether the estimate of entropy production can be improved any further using the realization that the observed dynamics must come from an underlying jump network. Might it even be possible to find an underlying model for the observed statistics? 

\subsection{Unicyclic networks} \label{sec:unicylcic_networks}
Networks with only one cycle are special because here time-series irreversibility already captures the total entropy production. Taking~\eref{eq:DKL_per_step} and splitting out the contribution from direct jumps gives:
\begin{equation}
    \hspace*{-1cm}\Delta\Sigma_\mathrm{DKL} = \left( a_{12} p_2 - a_{21} p_1 \right)\ln\frac{a_{12}}{a_{21}} + \sum\limits_{n=2}^\infty \left[ p_{12}(n)\, p_2 - p_{21}(n)\, p_1 \right]\ln\frac{p_{12}(n)}{p_{21}(n)}\,. \label{eq:Sigma_DKL_split_out_direct}
\end{equation}

\begin{figure}[htpb]
    \centering
    \includegraphics[width = 0.25 \linewidth]{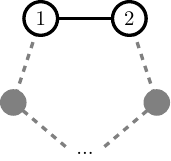}
    \caption{Topology of unicyclic network. States $1$ and $2$ are observed. All other states are hidden. A cyclic current flows over the edge $1-2$ and completes the cycle through the hidden part of the network.}
    \label{fig:unicyclic_network}
\end{figure}

Let there be $m$ hidden states. In the following it is assumed that the hidden part of the network is not disjoint and the visible states are directly connected such that a cycle that starts by a transition $1\to 2$ needs to be completed by transitioning through the hidden part of the network (see figure~\ref{fig:unicyclic_network}). Because the network is unicyclic and the numbering of the hidden states is arbitrary, they can be arranged such that the hidden matrix of transition probabilities $\bH$ is tri-diagonal. Then, the ratio of jump probabilites can be written as
\begin{equation}
    \frac{p_{12}(n)}{p_{21}(n)} = \frac{a_{1,m+2}\,\left[
    \bH^{n-2}\right]_{m1} a_{32}}{a_{m+2,1}\,\left[
    \bH^{n-2}\right]_{1m} a_{23}}\,.
\end{equation}

Now, powers of tri-diagonal matrices have a special property (see~\ref{sec:appendixA}), which permits simplification of this ratio to
\begin{eqnarray}
\hspace{-10mm}\frac{p_{12}(n)}{p_{21}(n)} &= \frac{a_{1,m+2}\, h_{m,m-1} \,...\, h_{32}\, h_{21} \,a_{32}}{a_{m+2,1}\, h_{m-1,m}\,...\,h_{23}\,h_{12}\,a_{23}} \label{eq:unicyclic_ratio_jumps_1} = \frac{p_{12}(m+1)}{p_{21}(m+1)}\, \quad \mathrm{for}\; n > m\,. \label{eq:unicyclic_ratio_jumps_2}
\end{eqnarray}
That is, the ratio of jump probabilities for $n$ time steps does not exist for $n \leq m$ because the hidden part of the cycle has more than $n$ states, and is independent of $n$ for $n > m$. In principle, this relation allows one to deduce whether a masked jump network is unicyclic from observations of the jump probabilities between the visible states alone. 

The proportion of time a certain visible state is visited obeys an equation similar to the Master equation:
\begin{eqnarray}
    \pi_2 = \sum\limits_{n=1}^\infty \left[ p_{21}(n)\, \pi_1 + p_{22}(n) \, \pi_2 \right]\,,
\end{eqnarray}
where $\pi_1$ and $\pi_2$ are given by~\eref{eq:steady-state_probs_cond}. Normalization of the jump probabilities, $1~=~\sum_{n=1}^\infty \left[p_{12}(n) + p_{22}(n)\right] \Leftrightarrow \sum_{n=1}^\infty p_{22}(n) = 1 - \sum_{n=1}^\infty p_{21}(n)$, then implies:
\begin{eqnarray}
    0 &= \sum\limits_{n=1}^\infty \left[ p_{21}(n)\, \pi_1 - p_{12}(n) \, \pi_2 \right]\\
    &= \sum\limits_{n=1}^\infty \left[ p_{21}(n)\, p_1 - p_{12}(n) \, p_2 \right]\,,
\end{eqnarray}
where the second line follows from~\eref{eq:steady-state_probs_cond}. Splitting out the direct jumps, for which $p_{ij}(1) = a_{ij}$, then yields
\begin{equation}
    a_{21}\,p_1 - a_{12}\, p_2 = - \sum\limits_{n=2}^\infty \left[ p_{21}(n)\, p_1 - p_{12}(n) \, p_2 \right]\,. \label{eq:master_eq_split_out_direct}
\end{equation}

Finally, inserting~\eref{eq:unicyclic_ratio_jumps_1}~and~\eref{eq:master_eq_split_out_direct} into~\eref{eq:Sigma_DKL_split_out_direct} and using $h_{ij} = a_{i+2,j+2}$ [see~\eref{eq:block_matrix}] yields
\begin{eqnarray}
    \Delta\Sigma_\mathrm{DKL} &= \left( a_{12}\,p_2 - a_{21}\, p_1 \right)\ln\frac{a_{12}\,a_{23}...a_{m+2,1}}{a_{1,m+2}...a_{32}\,a_{21}}\\
    &= \Delta\Sigma\,,
\end{eqnarray}
where the last equality follows from the Hill-Schnakenberg decomposition of entropy production as flux times affinity for one cycle~\cite{Schnakenberg1976,Hill1989}.

Thus, masked unicyclic networks are special as one can recover the complete entropy production from partial observation and computing the time-series irreversibility. Bisker \emph{et al}.~\cite{Bisker2017} found a similar behavior for one of the partial entropy productions for unicyclic networks.

\section{Entropy production from fitting a four-state network to observable statistics} \label{sec:fitting}
As shown in section~\ref{sec:time_series_irreversibility}, the statistics of the observable trajectories are completely determined by the jump probabilities $p_{ij}(n)$ and the occupation probabilities $p_1$ and $p_2$ of the visible states. The question is thus how one can fit a hidden network to these observables. In the following I assume that the correct number of hidden states is known and can be used to constrain the possible underlying transition matrices. Note that the detailed `topology' of the network is not known, i.e., one does not know whether certain transition rates are zero (as is the case for the example network in section~\ref{sec:example_network}).

The elements in the first sub-matrix in~\eref{eq:block_matrix} can be observed directly. Moreover, each column of the transition matrix must sum to one, which is used to re-parametrize the sub-matrix $\bB$
\begin{eqnarray}
    \bB &= \left( \begin{array}{cc}
        a_{31} & \tilde a_{H2} - a_{42}\\
        \tilde a_{H1} - a_{31} & a_{42}
        \end{array} \right)\,,
\end{eqnarray}
where $\tilde a_{Hi}$ is the coarse-grained transition probability from the visible state $i$ into the hidden part of the network. It can be directly inferred.

Next, the remaining parameters are determined from the first 8 jump probabilities $p_{ij}(2)$ and $p_{ij}(3)$ for jumps with length $2$ and $3$.
\begin{eqnarray}
    \mathbf{P}(2) &= \bC \, \bB\\
    \mathbf{P}(3) &= \bC \, {\bH}\, \bB \,,
\end{eqnarray}
where $P(n)$ is the matrix composed of the jump probabilities $\{p_{ij}(n)\}$.

In the following, it is assumed that the matrix $\mathbf{P}(2)$ is non-singular, which implies that both matrices $\bB$ and $\bC$ can be inverted. One then has:
\begin{equation}
    \bC =  \mathbf{P}(2) \,\bB^{-1} \label{eq:inference_C}
\end{equation}
and
\begin{eqnarray}
    \bH &=  \bC^{-1}\,\mathbf{P}(3) \,\bB^{-1}\\
    &= \bB \, \mathbf{P}^{-1}(2) \,\mathbf{P}(3)\, \bB^{-1}\,. \label{eq:H_in_terms_of_p}
\end{eqnarray}

This procedure yields a solution in terms of two free parameters $a_{31}$ and $a_{42}$. The only constraint that has not been used at this point is that the remaining two columns of the transition matrix must sum to one. However, the above procedure already satisfies this constraint, which means that the solution has two free parameters.

Importantly, even though only the first two nontrivial jump probabilities have been used, the masked transition probabilities are completely determined, because as long as the matrices involved are non-singular, one has for $n\geq 3$ [with~\eref{eq:jump_probabilities}]:
\begin{eqnarray}
    \mathbf{P}(n) &= \bC \, \bH^{n-3} \, \bB \, \bB^{-1} \bH \, \bB\\
    &= \mathbf{P}(n-1)\,\bB^{-1} \bH\, \bB\,.
\end{eqnarray}
Particularly [from~\eref{eq:H_in_terms_of_p}]
\begin{equation}
    \bB^{-1} \bH\, \bB  = \mathbf{P}^{-1}(2)\,\mathbf{P}(3)\,.
\end{equation}
Therefore, if $\mathbf{P}(2)$ and $\mathbf{P}(3)$ are known, all other jump probabilities follow.

Depending on the choice of the free parameters $a_{31}$ and $a_{42}$, the above equations may yield invalid solutions, i.e., finding $a_{ij} < 0$ (equivalently $a_{ij} > 1$), incompatible with a transition probability. In practice, one can do a numerical parameter sweep for $0 \leq a_{31} \leq \tilde a_{H1}$ and $0 \leq a_{42} \leq \tilde a_{H2}$ and find the combinations that yield valid solutions.

Once all allowed solutions have been found, a range of estimated entropy productions can be calculated from~\eref{eq:entropy_prod}.

A similar procedure as shown above can be used when there are more hidden and/or observed states. Crucially, the equations~\eref{eq:jump_probabilities} relating the jump probabilites to the elements of the transition matrix are non-linear, making solving the inverse problem of finding transition matrices that generate the given jump probabilities a difficult problem. The procedure above works when there is an equal number of observed and hidden states and the matrices of jump probabilities can be inverted. Otherwise, a big system of nonlinear equations must be solved in terms of (possibly many) free parameters. Such an endeavour is likely helped by an Eigendecomposition of the hidden transition matrix $\bH$, which allows to fit its eigenvalues $\lambda_k$ from the jump probabilities as 
\begin{equation}
    p_{ij}(n) = \sum\limits_{k} \alpha_{ij}^{(k)} \lambda_{k}^{n-2}\quad \mathrm{for}\; n> 1\,. \label{eq:sum_exponentials}
\end{equation}

\subsection{Example}
\begin{figure}[ht]
    \centering
    \includegraphics[width = 1 \linewidth]{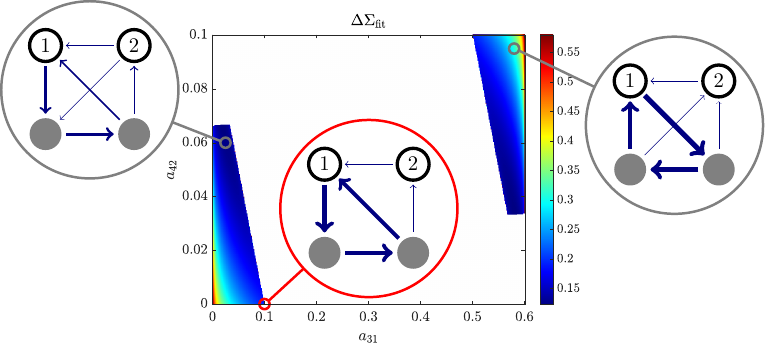}
    \caption{Estimated entropy production $\Delta\Sigma_\mathrm{fit}$ as a function of the free parameters $a_{31}$ and $a_{42}$. Jump probabilities are calculated using~\eref{eq:jump_probabilities} for the example transition matrix in~\eref{eq:example_transition_matrix} for $\Delta\mu=0$. Only valid solutions are plotted. The red circle marks the original transition probability matrix. The solution is symmetric with respect to replacing $a_{31}$ by $\tilde a_{H1} - a_{31}$ and $a_{42}$ by $\tilde a_{H2} - a_{42}$, since the numbering of the hidden states is arbitrary. Insets show the net steady-state probability currents produced by two example solutions (gray) and the original network (red).}
    \label{fig:EP_est_free_parameters}
\end{figure}

Figure~\ref{fig:EP_est_free_parameters} displays the entropy productions $\Delta\Sigma_\mathrm{fit}(a_{31},a_{42})$ resulting from the above fitting procedure for the observable jump probabilities calculated from the transition matrix in~\eref{eq:example_transition_matrix} for $\Delta\mu=0$ and all allowed combinations of $a_{31}$ and $a_{42}$. The solutions are symmetric with respect to replacing $a_{31}$ by $\tilde a_{H1} - a_{31}$ and $a_{42}$ by $\tilde a_{H2} - a_{42}$, which reflects the arbitrariness in the numbering of hidden states.

Notice that a wide range of entropy productions are compatible with the observed statistics of jumps. In fact, choosing ever-finer grids for $a_{31}$ and $a_{42}$ reveals that the highest allowed $\Delta\Sigma_\mathrm{fit}$ grows seemingly unbounded. This effect can be traced back to the possibility of having an almost absolutely irreversible transition along one edge $i\to j$ which contributes to the total entropy production with a huge edge-affinity $\ln a_{ji}/a_{ij}$. It is therefore prudent to take the lowest possible entropy production as an improved lower bound on the real entropy production:
\begin{equation}
    \Delta\Sigma^\mathrm{min}_\mathrm{fit} := \min_{a_{31},a_{42}} \Delta\Sigma_\mathrm{fit}(a_{31},a_{42})
\end{equation}
The real underlying entropy production is marked in red in figure~\ref{fig:EP_est_free_parameters}. Since the underlying transition rates satisfy $a_{2,4} = a_{4,2} =0$, they saturate the constraints on the allowed solutions. Therefore, the original network lies on the edge of the region of allowed solutions.

\begin{figure}[ht]
    \centering
    \includegraphics[width = 0.6 \linewidth]{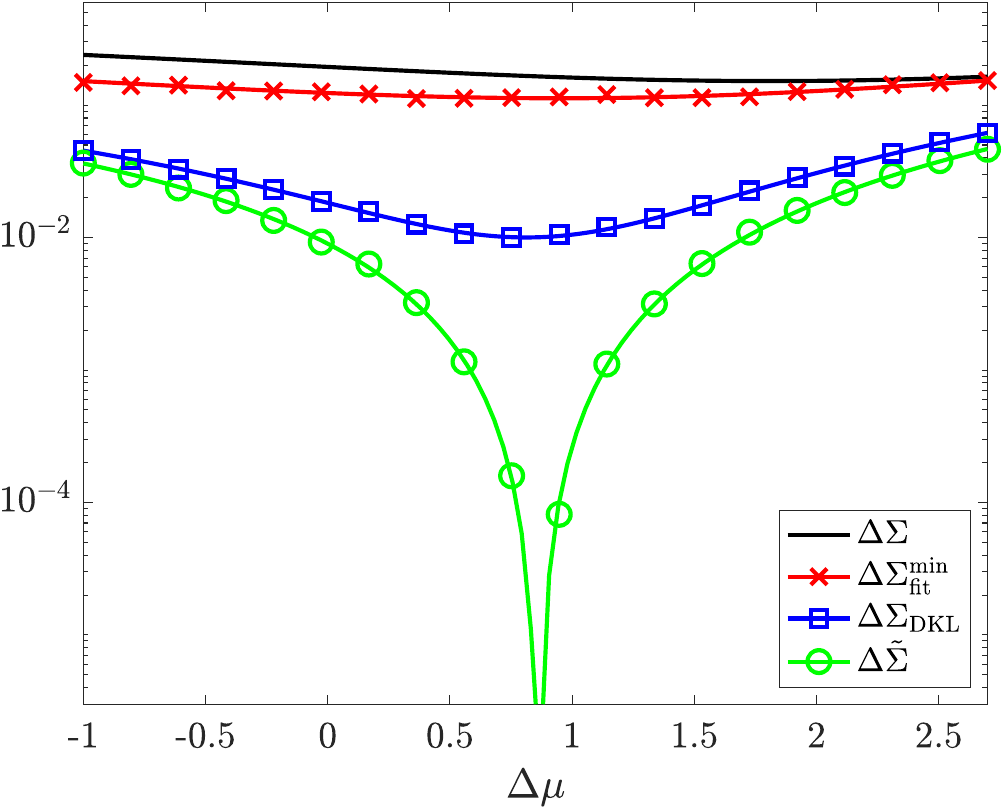}
    \caption{Real entropy production $\Delta \Sigma$~\eref{eq:entropy_prod} per time step, coarse-grained entropy production $\Delta\tilde\Sigma$~\eref{eq:coarse-grained_EP} resulting from mapping to a Markov model, observable time-series irreversibility $\Delta\Sigma_\mathrm{DKL}$~\eref{eq:DKL_per_step}, and estimated entropy production $\Delta\Sigma_\mathrm{fit}^\mathrm{min}$ from the fitting procedure, as a function of $\Delta\mu$. Solid lines are the entropy productions using analytical calculations of the transition and jump probabilities. Symbols are estimations based on transition and jump probabilities from a simulated trajectory of length $T=10^6$.}
    \label{fig:example_EP_2}
\end{figure}

Returning to the example in section~\ref{sec:example_network}, one can repeat the inference procedure for all values of $\Delta\mu$ and add a plot of the new lower bound to the curves in figure~\ref{fig:example_EP_1}, which is shown in figure~\ref{fig:example_EP_2}. It is evident that the entropy production inferred from fitting the jump probabilities does much better at capturing the full dissipation than the other methods.

\subsection{Random hidden two-state networks}
\begin{figure}[htpb]
    \centering
    \includegraphics[width = 0.6 \linewidth]{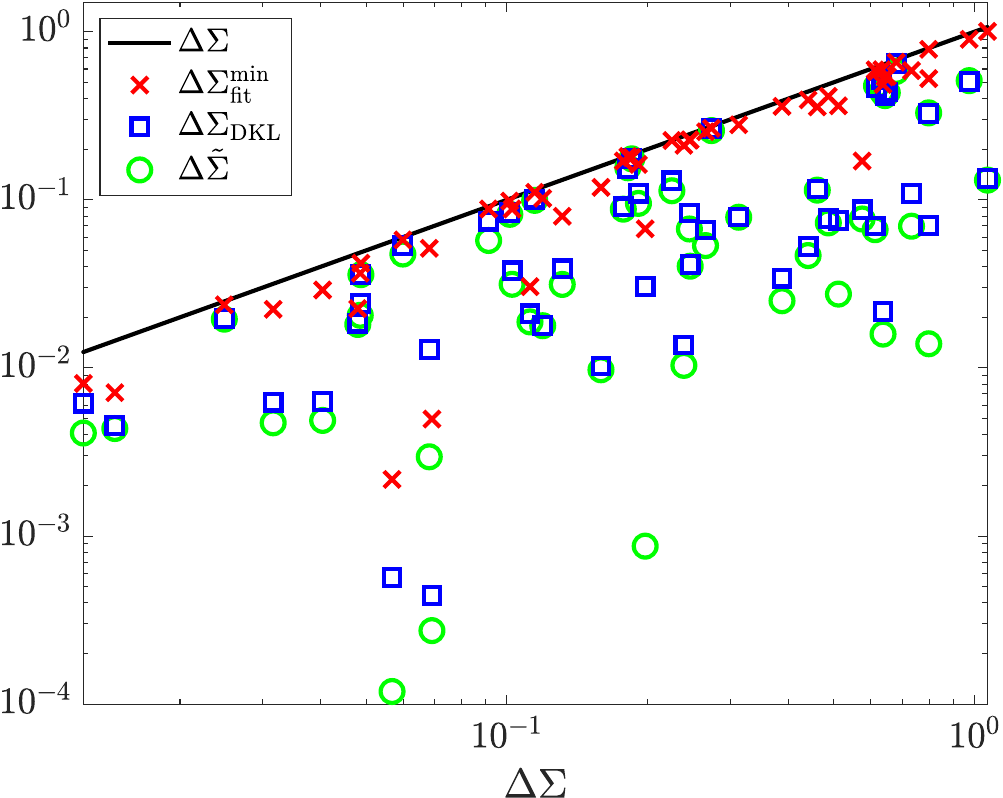}
    \caption{Comparison of real entropy production $\Delta \Sigma$~\eref{eq:entropy_prod} per time step with coarse-grained entropy production $\Delta\tilde\Sigma$~\eref{eq:coarse-grained_EP}, time-series irreversibility $\Delta\Sigma_\mathrm{DKL}$~\eref{eq:DKL_per_step}, and estimated entropy production $\Delta\Sigma_\mathrm{fit}^\mathrm{min}$ from the fitting procedure, for 50 randomly generated, fully connected transition matrices with two visible and two hidden states.}
    \label{fig:random_networks_EP}
\end{figure}

The inference procedure described above can also be applied to randomly generated transition matrices. For this, fully connected transition probability matrices $A$ were randomly generated. Next, the resulting coarse-grained steady-state probabilities and transition probabilities were calculated from~\eref{eq:coarse-grained_steady_state_prob}~and~\eref{eq:calc_coarse-grained_matrix}, respectively. These were used in~\eref{eq:coarse-grained_EP} to calculate the coarse-grained entropy production $\Delta\tilde\Sigma$ per time step. Next, the jump probabilities were calculated using~\eref{eq:jump_probabilities} and inserted into~\eref{eq:DKL_per_step} to calculate the time-irreversibility $\Delta\Sigma_\mathrm{DKL}$ per time step. Finally,~\eref{eq:inference_C}~and~\eref{eq:H_in_terms_of_p} were used to determine all possible transition matrices compatible with the jump probabilities. Of those, the one with lowest entropy production per time step $\Delta\Sigma_\mathrm{fit}^\mathrm{min}$ was chosen. Figure~\ref{fig:random_networks_EP} compares these irreversibility measures.

Again, the estimated entropy production $\Delta\Sigma_\mathrm{fit}^\mathrm{min}$ outperforms the other irreversibility measures. One finds the following hierarchy:
\begin{equation}
    \Delta\Sigma \quad\geq\quad \Delta\Sigma_\mathrm{fit}^\mathrm{min} \quad\geq\quad \Delta\Sigma_\mathrm{DKL} \quad\geq\quad \Delta\tilde\Sigma\,.
\end{equation}

\section{Discussion}
%% What can be generalized from results?
The inference strategy outlined here represents in many ways a `best-case' scenario because it is assumed that all of the observable statistics are available and determined to a high accuracy. The three estimators of the real entropy production I compared in this paper differ in how and in what way they use the observable data. The coarse-grained entropy production only uses a subset of it. Consequently, it produces the weakest bound on the real entropy production. Both the time-series irreversibility as well as the estimated entropy production from the fitting procedure in principle use all of the observable data. However, measuring time-series irreversibility is completely model-independent. In contrast, the fitting procedure described here requires that the number of hidden states is known which constrains the space of possible models that can generate the observable data. Consequently, it produces the tightest bound on the real entropy production.

This paper shows that in general even if perfect statistics about the observable jump trajectories are available, there is still some freedom in choosing model parameters such that one cannot infer a definitive entropy production.

%% What are unexpected results?
Unicyclic jump networks are an exception. As shown in sec.~\ref{sec:unicylcic_networks}, the observable jump probabilities allow the inference of the real entropy production and give a way to `spot' partially observed unicyclic networks from unique signatures in the observable data.

%% What are methodological limitations?
When trying the fitting procedure in practice, care must be taken to actually use all jump probabilities. While it is in principle possible to infer the hidden transition probability matrix from just the first two nontrivial jump probabilities, as shown here, this method is prone to errors. Instead, all jump probabilities for which good statistics are available should be combined in a similar way as in section~\ref{sec:fitting} to produce an estimate for the transition probability matrix.  

A limitation of the fitting procedure is that significant data are needed to generate good statistics for the transition probabilities. The problem is that not necessarily all of the in-principle \emph{observable} data are actually \emph{available} in a given scenario. 

%% How do own findings relate to other findings?
In practice, when underlying networks are complicated and good statistics rare, measuring time-series irreversibility~\cite{Martinez2019} might be the serviceable option to estimate entropy production. Alternatively, other strategies might be used that rely on more readily available subsets of the observable data. Skinner and Dunkel~\cite{Skinner2021} recently proposed one such method that also incorporates the demand for consistency with an underlying process. They show that a good estimator of entropy production can be obtained by counting the frequency of two successive jumps between coarse-grained states. From a general minimization procedure over possible generating models compatible with the statistics of this observable, they produce a simple estimator for entropy production that can outperform time-series irreversibility, despite requiring less data. While similar in spirit, the method presented here conserves the entire statistics of the observable trajectories, not just a subset of observables. It aims at capturing all of the entropy production by finding a true underlying model. This comes at the cost of requiring more data than a few jump rates.

More generally, demanding that observable data are consistent with an underlying physical process seems to offer the best direction for finding better estimators for entropy production of partially observed systems.

%% What are theoretical/practical applications?
Importantly, it is assumed here that the correct number of hidden states is known. In principle, one could try to infer this quantity by fitting a sum of exponentials [see~\eref{eq:sum_exponentials}] to the jump probabilities. This way, the number of terms needed corresponds to the number of eigenvalues of the hidden matrix. In practice, however, the statistics might be too poor for this to work. In any case, repeated (complex) eigenvalues are possible so that the hidden network could possibly be made larger while leaving intact the statistics of the jump probabilities. It is an interesting question for future studies if and how the bound found here degrades as more and more hidden states and cycles are added to the network. Could it degrade all the way to the lower bound given by time-series irreversibility, thus proving that this measure is the ultimate bound compatible with an underlying Markov process?

When the model generating the data is well known but its parameters need to be inferred, the method presented here is useful. It would also be possible to constrain the hidden matrix to only allow certain transitions, if the modelling supports such an assumption. If prior probabilities of specific models are known, a small extension of the fitting procedure could enable assigning probabilities to specific entropy productions based on the available data.

A further extension of the fitting method should also work when the hidden states are lumped into more than one hidden state and one observes `meso-dynamics' among several states. In analogy to Ref.~\cite{Martinez2019} the observable statistics are then completely determined by `semi-Markov' transition probabilities $p_{ijk}(n)$ encoding the probability to observe a jump form state $k$ to $j$, then wait in state $j$ for $n-2$ time steps, and subsequently jump from state $j$ to $i$.

Finally, it is expected that analogous results hold for continuous-time dynamics, with waiting time probability densities $p_{ijk}(t)$~\cite{Martinez2019} taking the place of the jump probabilities used here.

\section{Conclusion}
When inferring entropy production from statistics of partially observed dynamics, a number of different estimators are available. I have presented three: (1) Measuring the entropy production of an equivalent Markov process, (2) measuring time-series irreversibility, and (3) fitting an underlying physical process to the observable data and using its entropy production. While mapping to a Markov process produces the simplest estimator, it performs the worst, even failing to indicate any dissipation in the absence of observable currents. I have shown that measuring time-series irreversibility allows one to reconstruct the full entropy production when the underlying network is unicyclic. Using a partially observed jump process with four states, I have compared these estimators and find that estimating entropy production from fitting an underlying process produces the best estimate. When the underlying model is known and one only needs to fit its parameters, this method produces the tightest possible bound on the real entropy production.

\section*{Acknowledgments}
I thank Marcel Kahlen, Andreas Engel, Malte Behr, and Stefan Landmann for fruitful discussions and appreciate stimulating interactions with Juan M. R. Parrondo, Massimiliano Esposito, Matteo Polettini, Karel Proesmans, and John Bechhoefer. Furthermore, I thank David A. Sivak for helpful comments and a critical reading of the manuscript.

\section*{Data availability}
The code for generating the plots in this paper can be found in Ref.~\cite{code}.

\appendix
\section{Powers of tri-diagonal matrices}\label{sec:appendixA}
Consider an $m \times m$ tri-diagonal matrix $\bA$ with elements $a_{ij} \neq 0$ for $|i-j| \leq 1$ and $a_{ij} = 0$ for $|i-j| > 1$.

\subsection{Populating diagaonals} \label{sec:app_A_diagonals}
Each successive power of the matrix has the property to populate one further diagonal than the last. While the inner diagonals of the $n$-th power involve sums and products of matrix elements, the elements on the `outermost' diagonal are given by just one product of matrix elements.

Formally: For $m\geq i > j$ and $n \leq i-j$, the following holds:
\begin{equation}
    \left[ A^{n} \right]_{ij} = \cases{
    \prod_{k=j}^{i-1} a_{k+1,k} & for $n=i-j$\\
    0 & for $n < i-j$\,.
    }
\end{equation}

\subsubsection{Proof}
The statement is trivially true for $n=1$. Then, for $n> 1$, by induction:
\begin{eqnarray}
    \left[\bA^{n+1}\right]_{ij} &= \left[\bA^n\, \bA\right]_{ij}\\
    &= \Theta(j-1)\,\left[\bA^n\right]_{i,j-1}\,a_{j-1,j} + \left[\bA^n\right]_{i j}\,a_{j j} + \left[\bA^n\right]_{i,j+1}\,a_{j+1, j}\\
    &= \cases{
    a_{j+1,j}\, \prod_{k=j+1}^{i-1} a_{k+1,k} & for $n+1 = i-j$\\
    0 & for $n+1 < i-j$
    }\\
    &= \cases{
    \prod_{k=j}^{i-1} a_{k+1,k} & for $n+1 = i-j$\\
    0 & for $n+1 < i-j$
    }\,,\label{eq:lower_diagonal}
\end{eqnarray}
where
\begin{equation}
    \Theta(i) = \cases{
    1 & for $i \geq 1$\\
    0 & otherwise\,.
    }
\end{equation}

Similarly, for the 'upper' diagonals, one obtains
\begin{equation}
    \left[ A^{n} \right]_{ji} = \cases{
    \prod_{k=j}^{i-1} a_{k,k+1} & for $n=i-j$\\
    0 & for $n < i-j$\,.
    }\label{eq:upper_diagonal}
\end{equation}

\subsection{Ratio of elements}
The ratio of elements $(i,j)$ and $(j,i)$ of the $n$-th power of a tri-diagonal matrix is independent of $n$, if it exists.

Formally, for $m \geq i > j$ and $n \geq i-j$:
\begin{equation}
    \frac{\left[ \bA^n\right]_{ij}}{\left[ \bA^n\right]_{ji}} = \prod_{k=j}^{i-1} \frac{a_{k+1,k}}{a_{k,k+1}}\,. 
\end{equation}

\subsubsection{Proof}
From~\ref{sec:app_A_diagonals}, the statement is true for $n=i-j$. Then, using~\eref{eq:lower_diagonal}~and~\eref{eq:upper_diagonal}, by induction:
\begin{eqnarray}
    \hspace{-2cm}\left[\bA^{n+1}\right]_{ij} &= \left[\bA^{n} \, \bA\right]_{ij}\\
    &= \Theta(j-1)\,\left[\bA^n\right]_{i,j-1}\,a_{j-1,j} + \left[\bA^n\right]_{i j}\,a_{j j} + \left[\bA^n\right]_{i,j+1}\,a_{j+1, j} \label{eq:app_power_n_1_ij}\\
    \hspace{-2cm}\left[\bA^{n+1}\right]_{ji} &= \left[\bA \,\bA^{n} \right]_{ji}\\
    &= \Theta(j-1)\,a_{j,j-1}\,\left[ \bA^n \right]_{j-1,i} + a_{jj}\,\left[ \bA^n \right]_{ji} + a_{j,j+1}\, \left[ \bA^n \right]_{j+1,i}\\
    &= \Theta(j-1)\,a_{j,j-1}\prod_{k=j-1}^{i-1} \frac{a_{k,k+1}}{a_{k+1,k}}\,\left[\bA^n\right]_{i,j-1} + a_{j j}\prod_{k=j}^{i-1} \frac{a_{k,k+1}}{a_{k+1,k}} \, \left[\bA^n\right]_{ij} \nonumber\\
    &\qquad+ a_{j,j+1}\prod_{k=j+1}^{i-1} \frac{a_{k,k+1}}{a_{k+1,k}}\,\left[\bA^n\right]_{i,j+1}\\
    &= \Theta(j-1)\,a_{j-1,j}\prod_{k=j}^{i-1} \frac{a_{k,k+1}}{a_{k+1,k}}\,\left[\bA^n\right]_{i,j-1} + a_{j j}\prod_{k=j}^{i-1} \frac{a_{k,k+1}}{a_{k+1,k}} \, \left[\bA^n\right]_{ij} \nonumber\\
    &\qquad+ a_{j+1,j}\prod_{k=j}^{i-1} \frac{a_{k,k+1}}{a_{k+1,k}}\,\left[\bA^n\right]_{i,j+1}\\
    &= \prod_{k=j}^{i-1} \frac{a_{k,k+1}}{a_{k+1,k}}\,\left[\bA^{n+1}\right]_{ij}\,, \label{eq:app_power_n_1_ji}
\end{eqnarray}
where the last equality follows from~\eref{eq:app_power_n_1_ij}.

As a special case, one has for $n\geq m-1$ the relation:
\begin{equation}
    \frac{\left[ \bA^n\right]_{m1}}{\left[ \bA^n\right]_{1m}} = \prod_{k=1}^{m-1} \frac{a_{k+1,k}}{a_{k,k+1}}\,, \label{eq:app_ratio_powers_elements}
\end{equation}
which is used in the main text.

\section*{References}
\bibliographystyle{iopart-num}
\bibliography{references_masked_Markov}

\end{document}